\documentclass[%
 reprint,
 amsmath,amssymb,
prb,
]{revtex4-1}
\usepackage{graphicx}
\usepackage{color}
\usepackage{dcolumn} 
\usepackage{bm}      
\usepackage{amssymb}
\usepackage{amsfonts}
\usepackage{svg}
\usepackage{comment}
\usepackage{xr-hyper}
\usepackage[bookmarks=false,colorlinks,citecolor=red]{hyperref}
\usepackage{caption}
\usepackage{amsmath}
\usepackage{mhchem}
\usepackage{subcaption}
\usepackage[colorlinks,citecolor=red]{hyperref}
\include{MyCommand}

\newcommand{\COMMENTED}[1]{}

\definecolor{blue-violet}{rgb}{0.54, 0.17, 0.89}

\newcommand*{\citen}[1]{%
  \begingroup
    \romannumeral-`\x 
    \setcitestyle{numbers}%
    \cite{#1}%
  \endgroup   
}

\makeatletter
\newcommand*{\addFileDependency}[1]{
  \typeout{(#1)}
  \@addtofilelist{#1}
  \IfFileExists{#1}{}{\typeout{No file #1.}}
}
\makeatother

\newcommand*{\myexternaldocument}[1]{%
    \externaldocument{#1}%
    \addFileDependency{#1.tex}%
    \addFileDependency{#1.aux}%
}

\myexternaldocument{supplemental}

\newcolumntype{.}[1]{D{.}{.}{#1}}

\begin{document}

\title{Error Cancellation in Diffusion Monte Carlo Calculations of Surface Chemistry}

\author{Gopal R. Iyer}
\affiliation{Department of Chemistry, Brown University, Providence, RI 02912, USA}
\author{Brenda M. Rubenstein}
\email{brenda\_rubenstein@brown.edu}
\affiliation{Department of Chemistry, Brown University, Providence, RI 02912, USA}

\begin{abstract}
\textit{Abstract—}The accurate prediction of reaction mechanisms in heterogeneous (surface) catalysis is one of the central challenges in computational chemistry. Quantum Monte Carlo methods — Diffusion Monte Carlo (DMC), in particular — are being recognized as higher-accuracy, albeit more computationally expensive, alternatives to Density Functional Theory (DFT) for energy predictions of catalytic systems. A major computational bottleneck in the broader adoption of DMC for catalysis is the need to perform finite-size extrapolations by simulating increasingly large periodic cells (supercells) to eliminate many-body finite-size effects and obtain energies in the thermodynamic limit. Here, we show that it is possible to significantly reduce this computational cost by leveraging the cancellation of many-body finite-size errors that accompanies the evaluation of energy differences when calculating quantities like adsorption (binding) energies and mapping potential energy surfaces. We analyze the cancellation and convergence of many-body finite-size errors in two well-known adsorbate/slab systems, H$_2$O/LiH(001) and CO/Pt(111). Based on this analysis, we identify strategies for obtaining binding energies in the thermodynamic limit that optimally utilize error cancellation to balance accuracy and computational efficiency. Using one such strategy, we then predict the correct order of adsorption site preference on CO/Pt(111), a challenging problem for a wide-range of density functionals. Our accurate and inexpensive DMC calculations are found to unambiguously recover the top $>$ bridge $>$ hollow site order, in agreement with experimental observations. We proceed to use this DMC method to map the potential energy surface of CO hopping between Pt(111) adsorption sites. This reveals the existence of an L-shaped top–bridge–hollow diffusion trajectory characterized by energy barriers that provide an additional kinetic justification for experimental observations of CO/Pt(111) adsorption. Overall, this work demonstrates that it is routinely possible to achieve order-of-magnitude speedups and memory savings in DMC calculations by taking advantage of error cancellation in the calculation of energy differences that are ubiquitous in heterogeneous catalysis and surface chemistry more broadly.
\end{abstract}

\maketitle

\section{Introduction}

Solid-state heterogeneous catalysts have transformed the field of catalysis by enabling the selective catalysis of a wide-range of industrially important reactions, often at lower cost and in a more sustainable fashion than comparable homogeneous catalysts.\cite{norskov2014fundamental} Of particular significance has been the recent design of a variety of promising surface catalysts for the reduction of CO$_2$.\cite{co2_heterogeneous_catalysis} Heterogeneous catalysts achieve their unrivaled selectivity through the careful construction of their surfaces, which can be designed to complement specific reactants and thereby steer specific reactions.

While heterogeneous catalysts have historically been designed through a relatively time-consuming experimental guess-and-check process,\cite{schlogl_heterogeneous_catalysis} in more recent years, researchers have increasingly turned to simulation to predict how subtle surface modifications will impact reaction outcomes and accelerate the catalyst design process. The leading computational tool for modeling heterogeneous catalysis has long been Density Functional Theory (DFT),\cite{kohn_sham_dft} which can determine the potential energies of reactants, intermediates, and their products at relatively low costs enabling the treatment of solid surfaces and adsorbates that consist of hundreds of atoms.\cite{norskov_scheffler_toulhoat_2006,Hellman_Vincent_JPCB,Peterson_EnergyEnviron_2010,Jain_2013} Despite DFT's many triumphs modeling catalytic reactions,\cite{dft-review-norskov} it inherently struggles to capture electron correlation effects, which can play an essential role in the description of reactions involving covalent bond-breaking and formation, particularly at surfaces containing $d$- and $f$-block metals.\cite{Gaggioli_ACSCat,vogiatzis_multireference,gagliardi_multireference} The sometimes not-so-subtle errors that can arise from an incomplete description of electron correlation can lead to incorrect predictions of binding, reactivity, and other trends, leading researchers to design suboptimal catalysts, ultimately diminishing the computational catalysis program. A prime example of this was the initial, DFT-based modeling of the catalysis of CO on Pt(111), which originally predicted adsorption to occur at sites that were inconsistent with experimental observations,\cite{Feibelman_JPCB_2001} and which required a more careful use of more advanced functionals to correct.\cite{Olsen_JCP,Grinberg_JCP,Janthon_JPCC} Such examples — and the many examples yet to be identified for lack of comparison — motivate the development and exploration of other electronic structure methods for studying catalysts that can more accurately account for the effects of electron correlation. In this work, we therefore examine the ability of quantum Monte Carlo (QMC) methods to accurately model two important features that characterize heterogeneous catalytic reactions — adsorption (binding) energies and the potential energy surfaces associated with adsorbate diffusion.

Diffusion Monte Carlo (DMC)\cite{Ceperley_Alder_1980,Foulkes_RMP_2001,qmcpack_2020} — a real-space, ground-state QMC method — has long been viewed as the gold standard for modeling the ground state properties of correlated solids due to its ability to accurately describe electron correlation. Previous work has established QMC as a highly accurate method for describing the properties of correlated solids, including transition metal oxides,\cite{correlated_kent_benali_TiO2,correlated_kent_benali_Ti4O7,correlated_alfe_oxides,correlated_kent_Ca2CuO3} as well as doped and/or defective materials \cite{defect_alfe_mgo,defect_benali_hafnia,defect_benali_TiO2,defect_kent_benali_NiO,defect_kent_ZnO}. More recent studies have demonstrated similar accuracies for covalent solids, especially materials involving van der Waals interactions,\cite{covalent_benali_vdW,covalent_benali_SiC,covalent_kent_diamond,covalent_kent_graphite,covalent_benali_interlayer_graphene} encouraging its use in problems that require precise energy calculations, including heterogeneous catalysis.

Furthermore, DMC has a comparatively low, polynomial ($O(N^{3}-N^{4}$, where $N$ is the number of electrons) cost compared to Coupled Cluster Theory (CCSD(T)) and Full Configuration Interaction (FCI)–based methods.\cite{Foulkes_RMP_2001,Malone_PRB_2020,Motta_PRX_2017}

To determine bulk properties, the energies from many-body methods like DMC must furthermore be extrapolated to the infinite-size (thermodynamic) limit to correct for many-body finite-size effects, which conventionally requires costly calculations at a variety of increasingly large system sizes.\cite{drummond_finite_size_error,Lin_PRE_2001,Kwee_PRL_2008} Schemes for the control and correction of finite-size effects have been proposed in the past, including the method of special k-points\cite{needs_special_kpoints} and modifications to the way Coulomb interactions are treated.\cite{needs_coulomb, needs_coulomb_2} However, supercell extrapolation remains the most common choice (and consequently, a major computational bottleneck) for obtaining energies in the thermodynamic limit. As a result, DMC has largely been overlooked for modeling catalytic processes despite its promise of chemical accuracy, except for a handful of studies that have used DMC to model the adsorption of small molecules to surfaces. Known examples of these (at the time of writing of this paper) are the binding of H$_2$ on metal surfaces\cite{Karalti_PCCP,dmc_H2_on_Mg,hoggan_doblhoff-dier,Powell_JCP_2020}, CO on transition metal surfaces,\cite{hsing_CO_on_Pt, hoggan_co_pt_al,hoggan_copper_platinum} and H$_2$O on LiH(001)\cite{dmc_water_on_LiH} and hBN,\cite{AlHamdani_JCP_2014,Brandenburg2019,AlHamdani_JCP_2015} and O$_2$ on graphene.\cite{shin_graphene} While our focus here is on the DMC method, other many-body electronic structure theories such as Auxiliary-Field QMC\cite{shiwei_zhang_afqmc_1,shiwei_zhang_afqmc_2} and CCSD(T)\cite{berkelbach_cc_embedding} continue to be explored for surface science.

Calculations of binding energies and potential energy surfaces may be seen as part of a ubiquitous broader class of calculations that involve evaluating energy differences between two largely similar atomic configurations with small chemical or physical modifications, \textit{e.g.} the addition of an adsorbate to a slab or the displacement of an adsorbate from one adsorption site to another. 
In this work, we demonstrate that this subtraction of energies leads to robust cancellation of many-body finite-size errors. This enables DMC to be used to compute highly accurate adsorption energies and precisely map potential energy surfaces using relatively small supercells, dramatically reducing the computational cost of finite-size extrapolations or, in some cases, eliminating their necessity altogether.

\begin{figure}[t!]
        \begin{subfigure}[t]{4cm}
         \includegraphics[width=4cm]{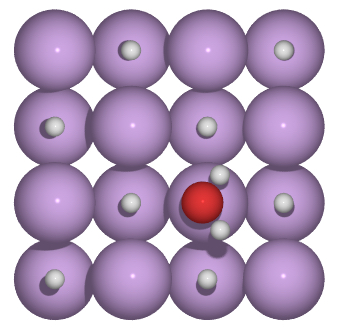}
        \caption{}
        \label{fig:H2O_on_LiH}
        \end{subfigure}
        ~
         \begin{subfigure}[t]{4cm}
         \includegraphics[width=4cm]{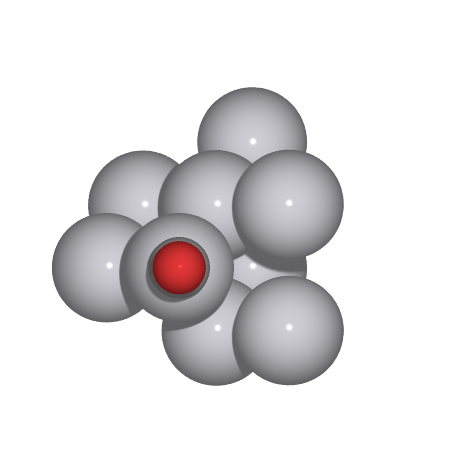}
        \caption{}
        \label{fig:CO_on_Pt111}
        \end{subfigure}
         \caption{Minimal unit cells of the two systems studied in this work: (a) \ce{H2O} on a two-layer LiH(001) slab and (b) CO on a three-layer Pt(111) slab. These specific systems and cell geometries were chosen since previous DMC studies on them\cite{dmc_water_on_LiH,hsing_CO_on_Pt} facilitate benchmarking.}
         \label{Models}
\end{figure}

Some previous works have implicitly recognized the role of error cancellation in DMC calculations of binding energies.\cite{dmc_H2_on_Mg, hsing_CO_on_Pt} In the following, we carry out a rigorous quantitative analysis of the cancellation and convergence of many-body finite-size errors for two illustrative slab/adsorbate systems. We begin by demonstrating the principle of error cancellation in the H$_2$O/LiH(001) system (Fig. \ref{fig:H2O_on_LiH}) and discuss its implications for accurate and inexpensive calculations of binding energetics. We then consider the CO/Pt(111) system (Fig. \ref{fig:CO_on_Pt111}), which is of greater practical interest and shows less trivial error cancellation behavior. Based on the observed error cancellation trend in this system, we compare the effectiveness of different finite-size extrapolation strategies to identify — in a generally applicable manner — the ones that optimally balance accuracy and computational efficiency. Subsequently, we apply one such optimally balanced extrapolation strategy to address the long-standing issue of determining the correct adsorption site preference order in CO/Pt(111). Finally, we use DMC to map the potential energy surface associated with the diffusion of the CO adsorbate between Pt(111) adsorption sites to obtain kinetic insights into experimental observations of preferred adsorption configurations. We conclude with a brief discussion of the computational gains made possible by properly leveraging error cancellation and reflect on its role in increasing the practicality of QMC simulations for a wider variety of problems in solid-state chemistry.

\section{Methods \label{methods}}

\subsection{Density Functional Theory and Nudged Elastic Band Calculations \label{dft}}

DFT calculations were performed for benchmarking as well as for the generation of the single-particle orbitals used to construct trial wave functions in subsequent Variational (VMC) and Diffusion Monte Carlo calculations. All DFT calculations were performed with Quantum ESPRESSO\cite{espresso} using the PBE functional.\cite{pbe_functional} Converged values have been chosen for the kinetic energy cutoff and the k-point grid. The details of the minimal unit cell geometries and DFT parameters used for each system are presented in Table \ref{tab:dft_details}. All DFT and QMC calculations were performed using Troullier–Martins pseudopotentials.\cite{troullier-martins-pseudopotential}

A climbing-image nudged elastic band (CI-NEB)\cite{cineb} calculation was performed to determine the lowest energy pathway for the diffusion of the CO adsorbate from the top to the fcc hollow site of Pt(111) (Fig. \ref{fig:site_hop_COPt}). For each geometry on the pathway, the energy was calculated using the same DFT settings mentioned above. A force convergence threshold of 0.05 eV/Å was used as the termination criterion for the NEB calculations.

\begin{table*}
\begin{tabular}{|c|c|c|c|}
     \hline
     \textbf{System} & \textbf{Slab geometry} & \textbf{k-point grid} &
     \textbf{$E_\textrm{cutoff}$ (Ry)} \\
     \hline
     \ce{H2O}/LiH(001) & 2$\times$2 (16 atoms/layer), 2 layers, $d_\textrm{ads–ads}$ = 8.522 Å & (10,10,1) & 350 \\
     \hline
     CO/Pt(111) & $\sqrt{3}\times\sqrt{3}$ (3 atoms/layer), 3 layers, $d_\textrm{ads–ads}$ = 4.830 Å & (4,4,1) & 260 \\
     \hline
\end{tabular}
\label{tab:dft_details}
\caption{\label{tab:dft_details} Minimal cell geometries and parameters used in the DFT calculations. $d_\textrm{ads–ads}$ is the distance between an adsorbate and its nearest periodic image.}
\end{table*}

\subsection{Diffusion Monte Carlo Calculations \label{dmc}}

All QMC calculations were performed using the QMCPACK suite.\cite{qmcpack_2020} For each supercell, Slater–Jastrow-type trial wave functions of the form $\Psi(\mathbf{R})=D^{\uparrow}D^{\downarrow}e^J$ were constructed using single-particle orbitals from the DFT calculations described above. (Here, $D^\uparrow$ and $D^\downarrow$ are Slater determinants composed of up- and down-spin single-particle orbitals, respectively, and $e^J$ is the corresponding Jastrow factor.) One- and two-body Jastrows were optimized using VMC simulations. Additionally, the single-particle orbitals for each system were encoded using the hybrid representation developed by Luo \textit{et al.}\cite{hybridrep_luo} This method has been shown to considerably improve memory efficiency while retaining or improving the accuracy of DMC energies. Following the optimization of the Slater–Jastrow wave functions, DMC calculations, each with a time-step of 0.01 a.u., were performed. 
Casula's T-move scheme for the variational evaluation of nonlocal pseudopotentials was employed.\cite{casula-t-moves} The number of DMC walkers was set to 2000 in each case, as this proved adequate to avoid walker population bias.

The same relaxed structures were used in both the DFT and QMC simulations.  Twist-averaged boundary conditions were used in the DMC calculations to correct for \textit{one-body} finite-size effects. To study \textit{many-body} finite-size effects, we performed DMC calculations on 1$\times$1, 1$\times$2, 1$\times$3, 2$\times$2, and 3$\times$3 tiled supercells of each system. 
In order to isolate many-body finite-size effects from one-body effects, we have normalized the twist-grid density in each direction by the inverse of the supercell tiling in that direction. For each of the systems studied herein, we use a (6,6,1) twist-grid for the 1$\times$1 cell. This then corresponds to a (6,3,1) grid for the 1$\times$2 supercell, a (3,3,1) grid for the 2$\times$2 supercell, etc. This ensures that energy differences between different supercells (normalized to the minimal cell) are solely due to many-body finite-size effects.

While our primary focus here is on many-body finite-size effects, for completeness, we note that DMC is susceptible to other sources of error, the most significant of these being the fixed-node error. A number of strategies for curtailing this error have been proposed, including variations on the use of configuration interaction using a perturbative selection made iteratively (CIPSI),\cite{fixed_node_scemama, fixed_node_benali, fixed_node_caffarel} and birth–death algorithms that require no \textit{a priori} information about the nodal surface.\cite{alavi_birth_death} Some other sources of error in DMC include pseudopotential errors, which may be remedied by stochastic projection–based methods (T-moves)\cite{casula-t-moves} with size-consistent adjustments,\cite{casula_pseudopotential_error} and time-step errors, which are controlled by modifying the approximate Green's function\cite{umrigar_timestep} and T-move scheme.\cite{timestep_and_pseudopotential_errors}

\subsection{Binding Energy Calculations \label{be}}


Binding energies for CO on Pt(111) were calculated using the standard formula,
\begin{equation}
    E_b = E_\textrm{slab+adsorbate} - E_\textrm{slab} - E_\textrm{isolated\_adsorbate}.
    \label{eq:binding_energy_CO_Pt}
\end{equation}
It has been shown, however, that it is possible to achieve significant cancellation of DMC time-step errors in binding energy calculations\cite{size_consistency} using the following modification:
\begin{equation}
    E_b = E_\textrm{slab+adsorbate} - E_\textrm{slab+displaced\_adsorbate}.
    \label{eq:binding_energy_H2O_LiH}
\end{equation}
Here, $E_\textrm{slab+displaced\_adsorbate}$ refers to the energy of the slab/adsorbate system with the adsorbate sufficiently displaced from the slab in the vertical direction so that the interactions between them are negligible. This alternative definition of binding energy has been used for the H$_2$O/LiH(001) system since it displayed better agreement with DMC benchmarks in the literature.\cite{dmc_water_on_LiH}

\subsection{Finite Size Extrapolations \label{fs}} 
It has been shown that for two-dimensional periodic systems, DMC energies evaluated using the Ewald interaction (as we have used here) scale as $O(N^{-5/4})$.\cite{drummond_finite_size_error} Using this scaling relation, the energy of a 2D system can be extrapolated to the thermodynamic limit using the energies of two or more supercells constructed by tiling a minimal unit cell. 

To quantify the finite-size error of a given cell, we define it as follows:\cite{hoggan_doblhoff-dier}
\begin{equation}
    \Delta \epsilon(N) = |E_{\infty} - E_N|,
    \label{eq:definition_finite_size_error}
\end{equation}
where $E_\infty$ is the extrapolated energy of the infinite system and $E_N$ is the energy corresponding to the supercell with $N$ electrons.

\section{Results and Discussion}

\subsection{Demonstration of Error Cancellation in the H$_2$O/LiH(001) System}
\label{section:h2o_lih}
To illustrate the cancellation of many-body finite-size errors in adsorption calculations, we first analyze its occurrence in binding energy calculations of H$_2$O adsorbed on the LiH(001) surface, a system that has been used extensively for benchmarking electronic structure methods.\cite{dmc_water_on_LiH, mp2_lih, fciqmc_lih} For reasons discussed later, the H$_2$O/LiH(001) system shows particularly pronounced cancellation of finite-size errors. However, we later also show how the same principles may be used to boost computational efficiency in a case where the degree of error cancellation is less-than-ideal, \textit{i.e.}, in the CO/Pt(111) system (Section \ref{section:co_pt}).

\begin{figure}
        \centering
         \includegraphics[width=9cm]{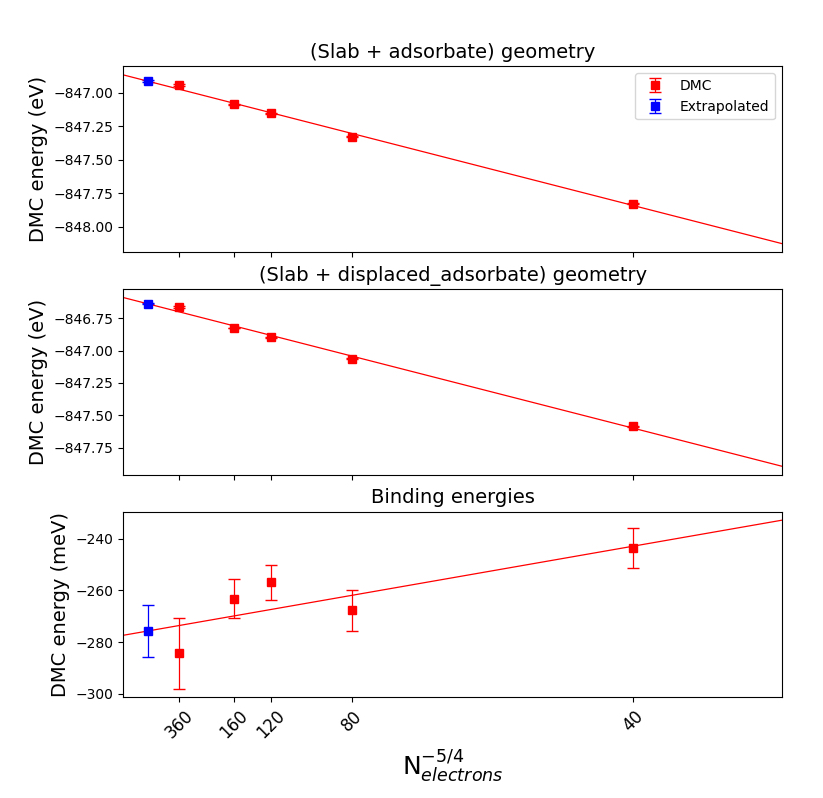}
         \caption{Finite-size extrapolation of \ce{H2O} adsorbed on a bilayer of LiH(001). Supercell sizes from right to left: 1$\times$1, 1$\times$2, 1$\times$3, 2$\times$2, and 3$\times$3.}
         \label{fig:plot_H2O_on_LiH001}
\end{figure}

The binding energy of the H$_2$O adsorbate on the LiH(001) slab is calculated by taking the difference between the total energies of the slab with a physisorbed adsorbate ($E_\textrm{slab+adsorbate}$) and the slab with a vertically displaced (non-interacting) adsorbate ($E_\textrm{slab+displaced\_adsorbate}$) as seen in Eq. \ref{eq:binding_energy_H2O_LiH}. For each of the energies computed, multiple DMC supercell calculations are carried out to perform finite-size extrapolations to the thermodynamic limit, as shown in Fig. \ref{fig:plot_H2O_on_LiH001}. This figure provides the first evidence of error cancellation: the energy range (y-axis) spanned by the different supercell and extrapolated energies is $\sim$1 eV for the two total energies (top two panels of Fig. \ref{fig:plot_H2O_on_LiH001}) but is limited to a few hundredths of an eV for the binding energy (bottom panel of Fig. \ref{fig:plot_H2O_on_LiH001}). This indicates that subtraction of the energies of two geometrically and chemically similar systems (slab+adsorbate and slab+displaced\_adsorbate) is accompanied by some degree of cancellation of many-body finite-size errors in their respective supercell energy calculations.

\begin{figure}
        \centering
         \includegraphics[width=9cm]{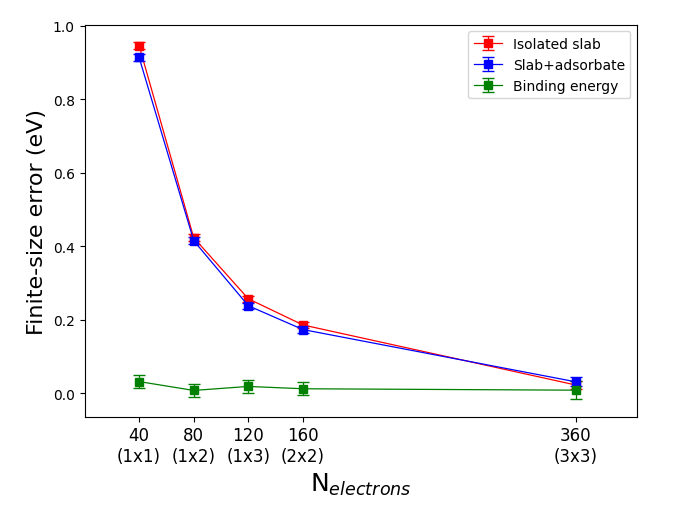}
         \caption{Comparison of finite-size errors for the isolated slab (with displaced adsorbate), the slab+adsorbate system, and the binding energy, as a function of system size for \ce{H2O}/LiH(001).}
         \label{fig:comparison_H2O_LiH}
\end{figure}

To quantify this cancellation of errors more concretely, we evaluate the many-body finite-size error associated with each supercell calculation with respect to the extrapolated energy according to Eq. \ref{eq:definition_finite_size_error}. Practically useful error cancellation can be said to occur when the finite-size error of the subtracted energy ($E_b$ here) is lower than that of the total energy, particularly for smaller supercells, while continuing to show the correct $O(N^{-5/4})$ scaling behavior. This is plotted for H$_2$O/LiH(001) in Fig. \ref{fig:comparison_H2O_LiH}. It can be clearly seen that error cancellation leads to a dramatic reduction — and almost immediate convergence — in the finite-size error of the binding energy. Thus, while a 3$\times$3 (or larger) supercell is needed to completely eliminate the finite-size error in the total energies, error cancellation allows the use of a 1$\times$1 or 1$\times$2 supercell to calculate highly accurate binding energies at a fraction of the computational cost. 

In any of these types of calculations, identifying the smallest supercell that shows a sufficient degree of error cancellation allows one to bypass the computationally demanding process of repeatedly running DMC calculations on progressively larger supercells of the same system in order to extrapolate to the thermodynamic limit. This is particularly relevant for mapping potential energy landscapes as doing so typically involves multiple calculations on slightly modified geometries which show similar finite-size error cancellation behavior. We explore the accurate and efficient mapping of potential energy landscapes facilitated by error cancellation further in the next section. We do so with the example of CO/Pt(111) adsorption, a theoretically and practically important system in heterogeneous catalysis. 

\subsection{Application to the CO/Pt(111) System}
\label{section:co_pt}

CO adsorbed on the Pt(111) surface is a well-known system in experimental and computational heterogeneous catalysis.\cite{experimental_co_pt,dft_co_pt} In this section, we discuss how error cancellation can be leveraged to probe this system with accurate and inexpensive DMC calculations. We begin by analyzing finite-size error cancellation in the binding energy. This information is then used to determine an optimal and generally applicable strategy for extrapolating binding energies to the thermodynamic limit. We then apply this strategy to map the CO/Pt(111) potential energy surface. In particular, we address the contentious issue\cite{Feibelman_JPCB_2001,Olsen_JCP,Grinberg_JCP,Janthon_JPCC} of identifying the correct order of CO adsorption site preference. We also attempt to determine the existence of kinetic barriers to the diffusion of the CO adsorbate between adsorption sites. These results are finally used to suggest a possible explanation for experimental observations of CO/Pt(111) binding.

\subsubsection{Analysis of error cancellation: Identifying an efficient strategy for finite-size extrapolation}
\label{section:2-point_extrapolation}
\begin{figure}
        \centering
         \includegraphics[width=9cm]{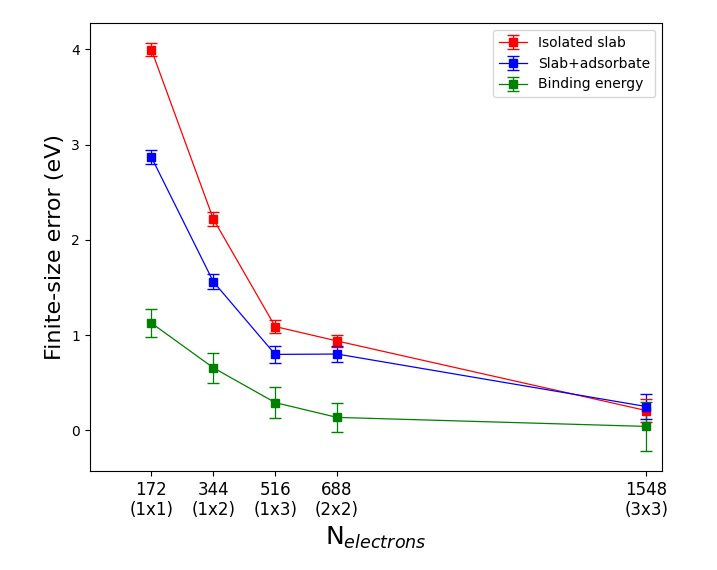}
         \caption{Comparison of finite-size errors for the isolated slab, the slab+adsorbate system, and the binding energy, as a function of system size for CO/Pt(111).}
         \label{fig:comparison_CO_Pt}
\end{figure}

Before utilizing error cancellation, it is worth checking whether it does indeed take place for the system being studied. Here, we analyze error cancellation in the binding energy of CO adsorbed at the top site of Pt(111) at a coverage of 1/3 monolayer (Fig. \ref{fig:CO_on_Pt111}). Similar to the H$_2$O/LiH(001) case, this is plotted for five supercells in Fig. \ref{fig:comparison_CO_Pt}. It can be seen that error cancellation visibly occurs for CO/Pt(111) as well. However, this cancellation does not occur as rapidly as in the case of H$_2$O/LiH(001). There are two possible reasons for this: (1) H$_2$O is physisorbed to LiH(001), while CO is chemisorbed to Pt(111), meaning that H$_2$O binds more weakly to LiH(001) than CO does to the Pt(111) surface. It is not unreasonable to expect physisorption and chemisorption to result in different error cancellation behavior because physisorption, almost by definition, results in more mild changes to the electronic structure upon binding. (2) The conventional formula used to calculate the CO/Pt(111) binding energy (Eq. \ref{eq:binding_energy_CO_Pt}) does not lend itself to error cancellation as easily as the formula used for H$_2$O/LiH(001) (Eq. \ref{eq:binding_energy_H2O_LiH}). This issue, too, is not always avoidable since Eq. \ref{eq:binding_energy_H2O_LiH} is not universally applicable. For example, the stoichiometry of molecules involved in surface modifications (such as the addition of a single O adatom) may not correspond to the stable molecular forms of those molecules (e.g., O$_2$). In such cases, Eq. \ref{eq:binding_energy_CO_Pt} is clearly the more practical choice.

\begin{figure}
        \centering
         \includegraphics[width=9cm]{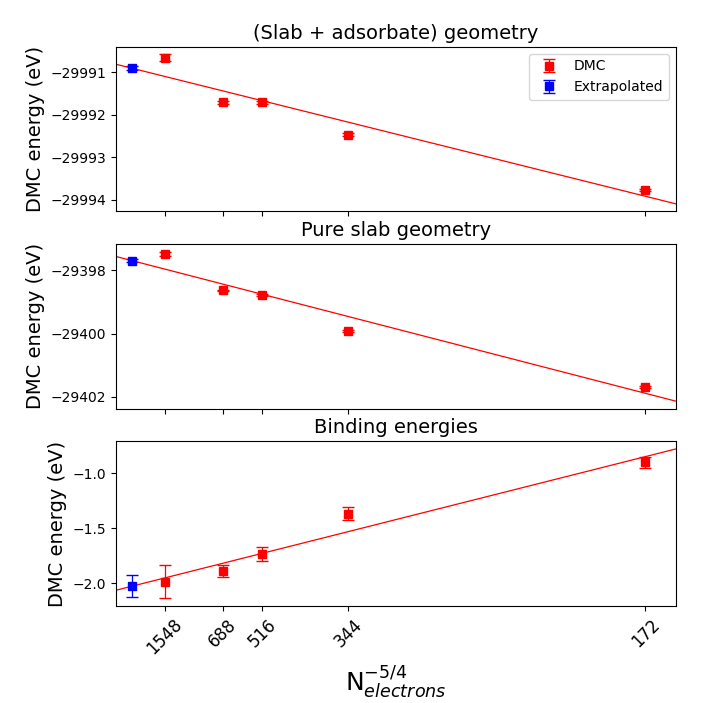}
         \caption{Finite-size extrapolation of CO adsorbed on the top-site of the Pt(111) slab at 1/3 monolayer coverage. Supercell sizes from right to left: 1$\times$1, 1$\times$2, 1$\times$3, 2$\times$2, and 3$\times$3.}
         \label{fig:plot_CO_on_Pt111}
\end{figure}

While error cancellation accelerates the convergence of finite-size errors on binding energies, it is difficult to determine \textit{a priori} which is the smallest supercell that will show adequately small errors. Therefore, it is important to have a general strategy for leveraging error cancellation even when it does not result in the immediate convergence of finite-size errors for a given system. For adsorption calculations, we can ask: What is the minimum amount of data needed to obtain reliable estimates of binding energies in the thermodynamic limit? 

In the past, the predicted $O(N^{-5/4})$ scaling behavior of the finite-size error in two-dimensional systems has, in some cases, been exploited by simply finding the equation of a line that passes through two supercell data points.\cite{hoggan_doblhoff-dier,Powell_JCP_2020} While this strategy promotes computational efficiency by relying on only two supercell calculations, it is important to consider whether the two supercell data points being used are sufficiently well-behaved to permit an accurate extrapolation. In the context of calculating quantities like binding energies, there are two aspects to this problem. First, one can ask whether error cancellation does in fact lead to better-behaved supercell binding energies allowing for more reliable extrapolations. More importantly, one can check how well any given extrapolation strategy agrees with other possible strategies using fewer, more, or the same number of supercell data points. We address these two issues in turn.

\begin{figure}
        \centering
         \includegraphics[width=9cm]{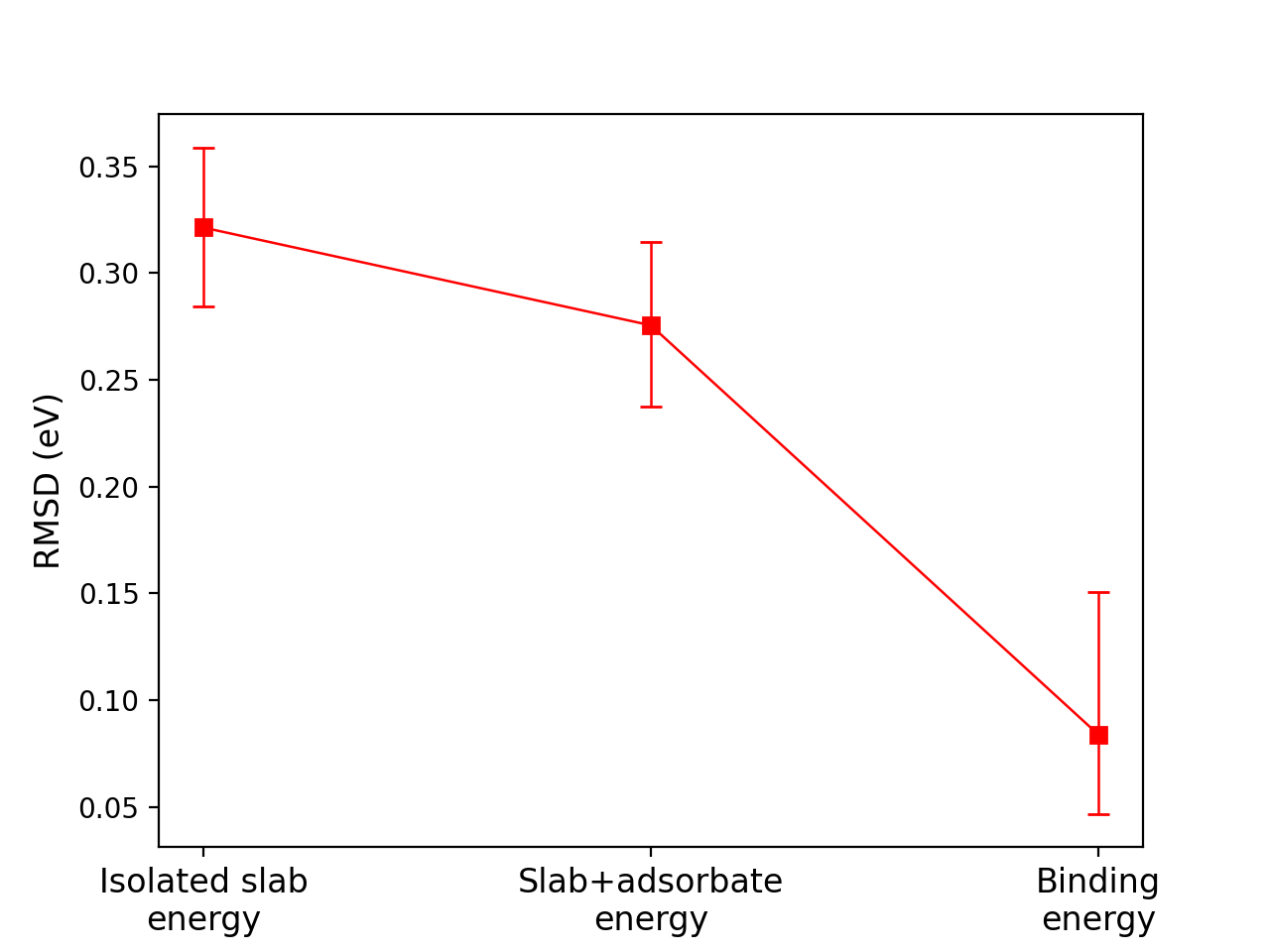}
         \caption{Comparison of the RMSDs from the finite-size extrapolation of absolute energies and additive interactions for CO/Pt(111).}
         \label{fig:rmsd_CO_Pt}
\end{figure}

If it can be shown that the fit in the bottom panel of Fig. \ref{fig:plot_CO_on_Pt111} is ``closer" than the fits in the two upper panels, the binding energy data points can be said to collectively show greater adherence to the $O(N^{-5/4})$ scaling law compared to the total energies due to error cancellation. To measure the closeness of the fit, we compute its root-mean-square deviation (RMSD). We compare this for finite-size extrapolations of the isolated Pt(111) slab energy, (top-)CO/Pt(111) energy, and the binding energy (Fig. \ref{fig:rmsd_CO_Pt}). The RMSD of the fit for the binding energy is clearly seen to be smaller than that of the isolated slab and CO/Pt(111) total energies. This verifies that, due to error cancellation, the binding energy data points are indeed more well-behaved and better suited to extrapolation than the total energies. (We also observe this trend for the H$_2$O/LiH(001) system, see Fig. S1 in the Supporting Information. The RMSD can, thus, be used as a metric to assess the reliability of error cancellation.) 

Since error cancellation leads to better quality supercell data points for extrapolating binding energies to the thermodynamic limit, one can anticipate that \textit{fewer} of these data points are needed to carry out the extrapolation reliably. We will now verify this numerically by calculating the binding energy of CO absorbed on the top site of Pt(111). As an example, we perform a 2-point extrapolation using the two smallest square supercells: 1$\times$1 and 2$\times$2. (This choice is merely illustrative and the same extrapolation can also be accurately performed using other supercell sizes, as explained below.) Before proceeding, we note that DMC predictions of binding energies can depend on a number of factors, particularly the DFT functional and pseudopotentials used to generate the trial wave function. With this caveat in mind, we predict that the binding energy of CO adsorbed at the top site of the Pt(111) surface is $-2.103 \pm 0.052$ eV. This is found to be in agreement with several dispersion-corrected and -uncorrected GGA and meta-GGA DFT functionals that predict binding energies in the range of $-1.8$ to $-2.4$ eV.\cite{Janthon_JPCC} It is encouraging to see that our DMC binding energy lies within the range of DFT values. 

\begin{figure}
        \centering
         \includegraphics[width=9cm]{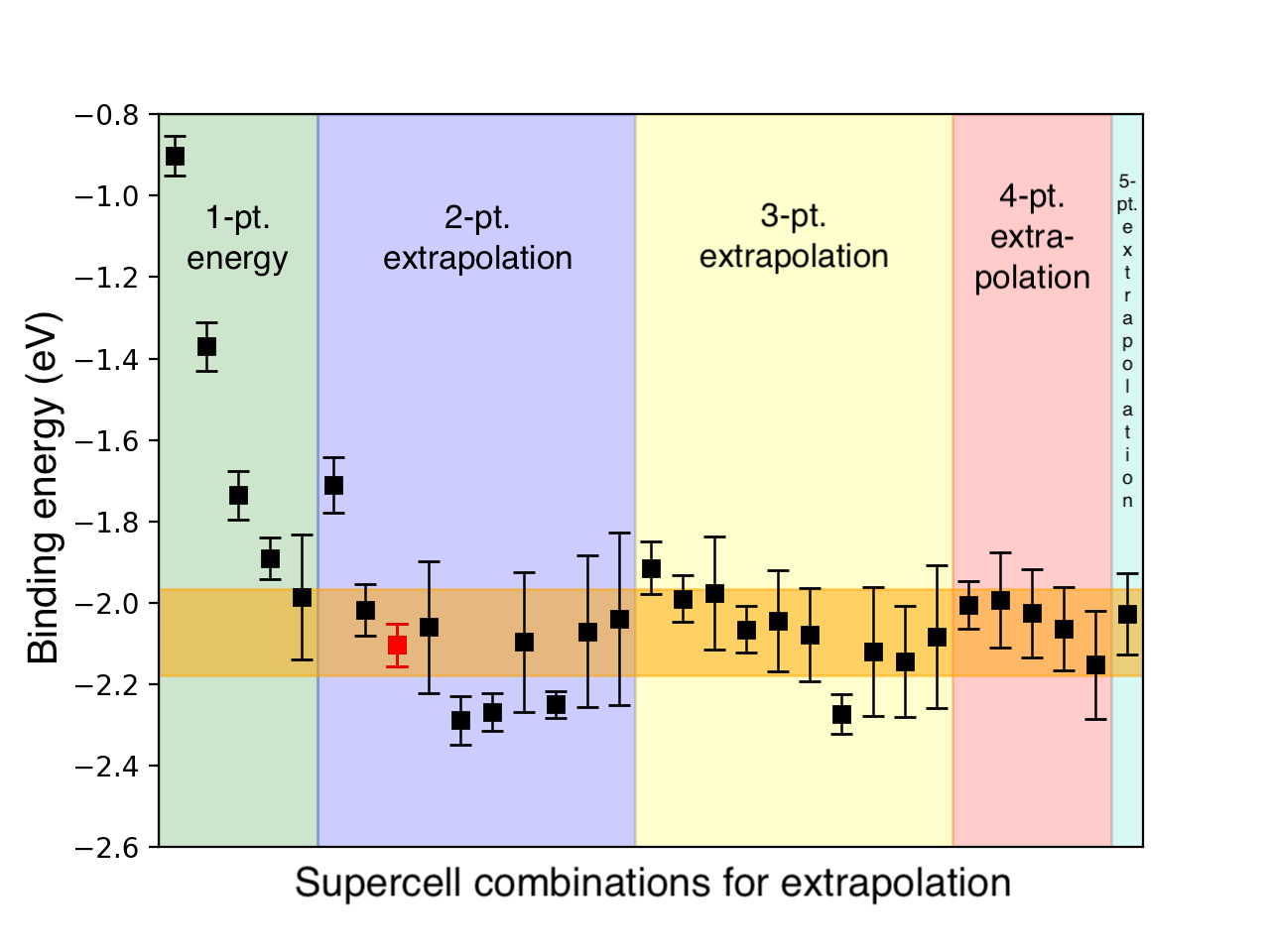}
         \caption{DMC binding energies calculated with one supercell, and extrapolated using 2, 3, 4, and 5 supercells. Supercell combinations for extrapolation are taken in order from the set \{(1$\times$1), (1$\times$2), (1$\times$3), (2$\times$2), (3$\times$3)\} and plotted from left to right. Horizontal orange shaded region: region of agreement between different possible extrapolations. Red point indicates energy extrapolated using 1$\times$1 and 2$\times$2 supercells.}
         \label{fig:internal_consistency_CO_Pt}
\end{figure}

In the preceding calculation, we have selected two of the five possible supercells, \{(1$\times$1),\ (1$\times$2),\ (1$\times$3),\ (2$\times$2),\ (3$\times$3)\}, to extrapolate binding energies. However, it is possible to perform the extrapolation in a total of 26 ways based on whether we use any 2, 3, 4, or all 5 of these supercell data points: ${5\choose2}+{5\choose3}+{5\choose4}+{5\choose5}=26$. It would, then, be informative to check whether our specific 2-point extrapolation (using the 1$\times$1 and 2$\times$2 supercells) is also internally consistent with all of the other possible ways of extrapolating DMC energies. In Fig. \ref{fig:internal_consistency_CO_Pt}, we plot the DMC binding energies obtained using individual supercells, and using each of these 26 possible extrapolations. Our goal is to determine how well any given extrapolation agrees with all of the others. To check this, we take the unweighted mean of all of the extrapolated energies and their error bars. The range of energies spanned by this mean indicates a region of global agreement among the different possible extrapolations (horizontal orange shaded region in Fig. \ref{fig:internal_consistency_CO_Pt}). It can be seen that 6 out of 10 2-point extrapolations, 9/10 3-point extrapolations, 5/5 4-point extrapolations, and 1/1 5-point extrapolation show overlapping error bars within the region of agreement. In particular, our 2-point extrapolation using the 1$\times$1 and 2$\times$2 supercells (marked in red in Fig. \ref{fig:internal_consistency_CO_Pt}) lies completely within this region. This is despite the fact that the individual supercell energy of neither the 1$\times$1 nor the 2$\times$2 cell overlaps with this region. We also see that some extrapolated energies \textit{do not} overlap with the region of agreement. Unsurprisingly, these correspond to using fewer and smaller supercells for extrapolation. It is worth mentioning that these are not necessarily ``bad" extrapolations. While they predict moderately different binding energies than the rest, they all lie comfortably within the range of numerous DFT-based predictions.\cite{Janthon_JPCC, Grinberg_JCP}

Furthermore, when mapping the potential energy surface of a system, absolute values of binding energy are only relevant insofar as they establish an energy baseline corresponding to a given configuration of atoms. Rather, one is primarily interested in the \textit{relative} binding energies of different atomic configurations with respect to some equilibrium geometry. As we will show in Section \ref{section:potential_energy_landscape}, it is possible to accurately resolve relative binding energies of nearby points on the potential energy surface in these systems by using only two supercells for DMC finite-size extrapolation. 

From the foregoing analysis, we see that the approach employed in previous works\cite{hoggan_doblhoff-dier,Powell_JCP_2020} of using two DMC data points for extrapolating binding energies is a quantitatively justified and computationally desirable strategy. In the discussion that follows, we continue to perform 2-point finite-size extrapolations using the 1$\times$1 and 2$\times$2 supercells.

\subsubsection{Mapping the CO/Pt(111) potential energy surface}
\label{section:potential_energy_landscape}
\begin{figure}
        \centering
         \includegraphics[width=9cm]{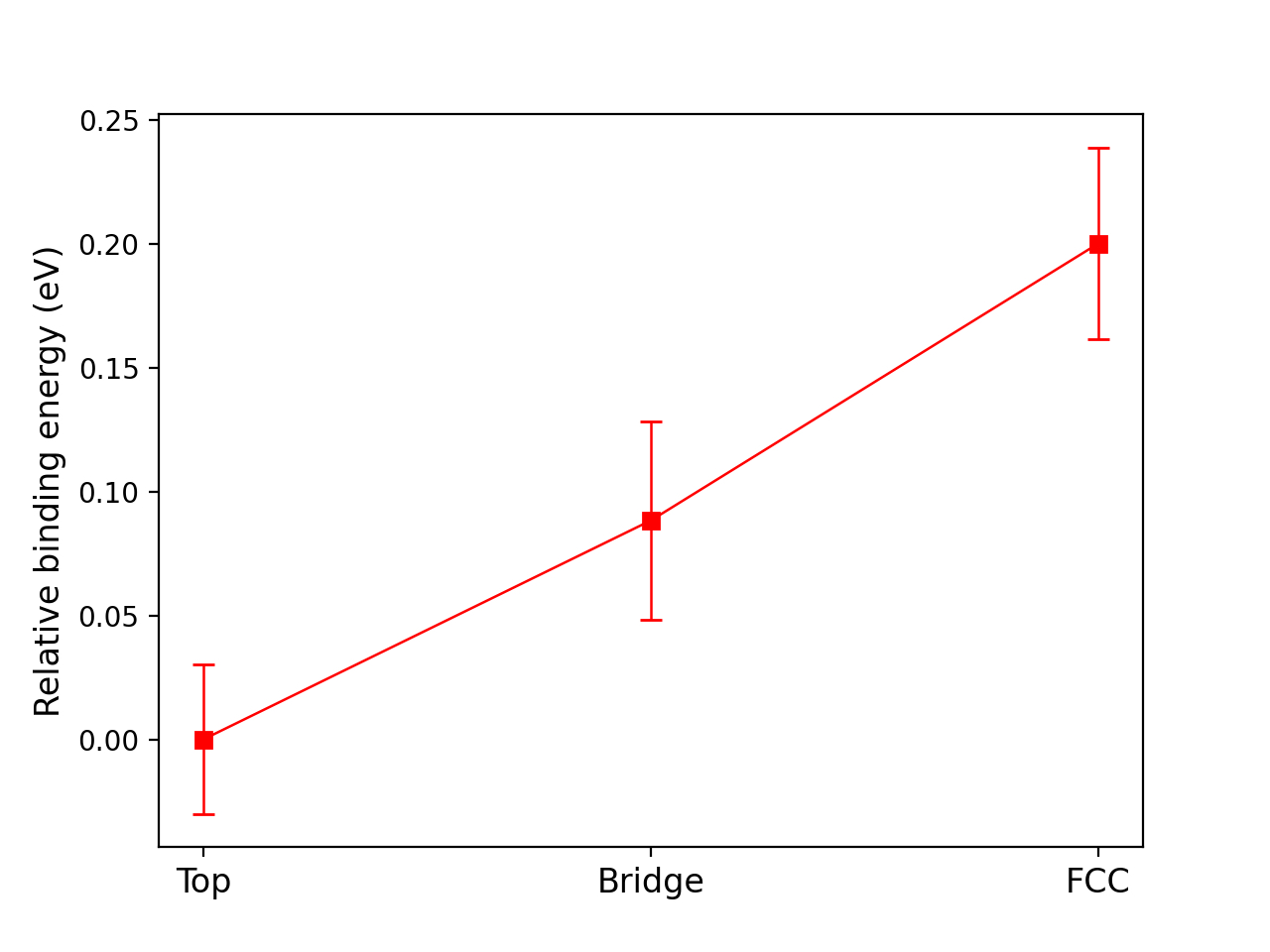}
         \caption{DMC binding energies of CO on Pt(111) adsorption sites relative to the top site, obtained using two supercell data points (1$\times$1 and 2$\times$2) for extrapolation.}
         \label{fig:site_pref_COPt}
\end{figure}

Next, we attempt to establish the order of adsorption site preference on CO/Pt(111). We are now concerned only with binding energies relative to a certain baseline (say, the top-site binding energy). The statistical error bars of the DMC binding energies can, then, be significantly reduced. This is achieved by noting that the DMC statistical errors associated with the isolated Pt(111) slab and the isolated CO molecule will remain exactly the same for every calculated binding energy (Eq. \ref{eq:binding_energy_CO_Pt}). These therefore cancel out when differences in binding energies are computed. This affords us the required statistical precision to clearly identify the order of adsorption energies as seen in Fig. \ref{fig:site_pref_COPt}. In agreement with experimental\cite{spectroscopy_co_pt_site_pref, stm_co_pt_site_pref} and some theoretical\cite{Olsen_JCP, mason_prb} studies, we correctly predict that the order of adsorption site preference is top $>$ bridge $>$ fcc hollow. We note that this result agrees qualitatively with that of Hsing \textit{et al.}\cite{hsing_CO_on_Pt} However, those authors predict that the top site is favored over the fcc hollow site by a much greater margin ($\sim$0.76 eV) than we do ($\sim$0.25 eV). This is the only other work we know of that uses DMC to study the CO/Pt(111) system. On the other hand, previous DFT studies have routinely shown that differences in binding energies between these sites is rarely greater than 0.3 eV.\cite{Feibelman_JPCB_2001, Grinberg_JCP, Olsen_JCP, mason_prb, Janthon_JPCC} We take this to be a confirmation of the accuracy of our calculations.

We now aim to test whether our 2-point–extrapolated DMC calculations can map the CO/Pt(111) potential energy surface at a finer level. This can be done by selecting points that lie closer to each other in molecular configuration space. In addition to validating the extrapolation method, this would also offer valuable insights into the kinetics of CO adsorption on Pt(111) at the level of a many-body theory. Here, we choose points along the diffusion trajectory from one adsorption site to another. Specifically, our goal is to map the potential energy surface for the diffusion of the CO molecule from the top to the fcc hollow site of Pt(111). In the past, there have been differing approaches to using QMC methods to map diffusion or reaction pathways. One approach involves calculating the energies of structures obtained by intentionally displacing the diffusing species (adsorbate) along a uniform grid of interpolated points between the initial and final positions.\cite{shin_graphene,kent_Li_graphite} Alternatively, one can rely on DFT-based NEB calculations\cite{dmc_H2_on_Mg} or semi-empirical methods like SRP-DFT\cite{hoggan_doblhoff-dier,Powell_JCP_2020} (intended to reproduce experimental barrier heights) to provide structures corresponding to a reactive trajectory which are then studied using QMC methods. Recently, it has also become possible to use QMC-level approaches to obtain optimized geometries and minimum energy pathways.\cite{vmc_reaction_pathway_1,vmc_reaction_pathway_2,vmc_reaction_pathway_3} However, this is accompanied by the increased computational cost of QMC methods. Here, we take the second approach (following the precedent of Ref. \citen{dmc_H2_on_Mg}), wherein we perform a DFT-NEB calculation to suggest points along the top–fcc hollow diffusion trajectory. We then perform DMC calculations on these points. This approach strikes a balance between accuracy (since it relies on physically meaningful structures suggested by the DFT-NEB trajectory) and computational efficiency (since it uses DFT instead of QMC to obtain the diffusive trajectory).

\begin{figure}
        \centering
         \includegraphics[width=9cm]{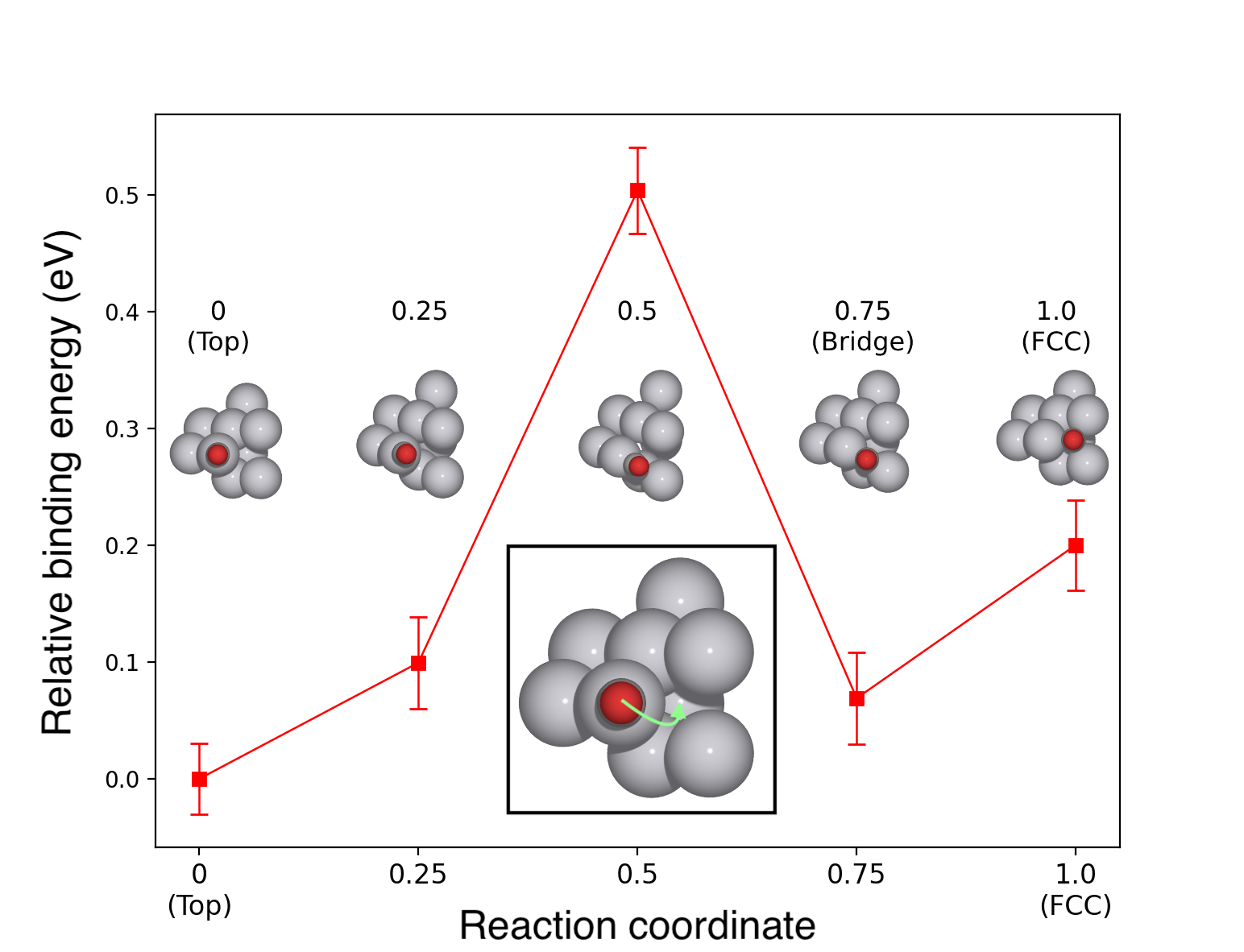}
         \caption{DMC energies (relative to the top site) of points on the path from the top to the fcc hollow site, obtained using two supercell data points (1$\times$1 and 2$\times$2) for extrapolation. The penultimate point is geometrically and energetically identical to the bridge site. Inset: L-shaped top–fcc hollow diffusion pathway suggested by the CI-NEB calculation (green arrow).}
         \label{fig:site_hop_COPt}
\end{figure}

In a seminal paper, Feibelman \textit{et al.}\cite{Feibelman_JPCB_2001} suggested that the issue of adsorption site preference in the CO/Pt(111) system may be partly explained by the existence of a hypothetical energy barrier that prevents the diffusion of the CO molecule to the 3-fold hollow sites on the metal surface. While past \textit{ab initio} studies have focused on identifying the correct order of adsorption site preference,\cite{Feibelman_JPCB_2001,Grinberg_JCP,Olsen_JCP,hsing_CO_on_Pt} ours is the only work we know of that attempts to map this CO/Pt(111) potential energy barrier using a many-body theory. Choosing the top site and the fcc hollow site as the initial and final points, respectively, we carry out a climbing-image nudged elastic band calculation to identify three intermediate points along the minimum energy pathway. Visual inspection of the penultimate (fourth) point in the overall trajectory reveals that it is geometrically identical to the bridge site (Fig. \ref{fig:site_hop_COPt}). This is further confirmed by their closely matching DMC energies. The NEB calculation, therefore, suggests an L-shaped trajectory for top–fcc hollow diffusion (inset: Fig. \ref{fig:site_hop_COPt}): the CO molecule first descends from the top to the bridge site, then falls into the fcc hollow site. The DMC energies for the points on this trajectory are shown in Fig. \ref{fig:site_hop_COPt}. Clearly, the top site is most favored for adsorption, and a significant energy barrier exists for diffusion to the fcc hollow site. It is interesting to note that a local minimum exists immediately adjacent to the fcc hollow site, which we have classified as the bridge site from the NEB trajectory. Mapping the potential energy surface in this way provides a more complete perspective on the experimentally observed adsorption configurations of CO on Pt(111). The top site is unambiguously the most stable; the bridge site acts as a local minimum during attempted top–fcc hollow diffusion, which explains the partial bridge-site adsorption seen in electron energy loss spectroscopy (EELS) experiments.\cite{experimental_co_pt} The fcc hollow site can only be attained by scaling \textit{two} energy barriers (top–bridge and bridge–hollow) during top–fcc hollow diffusion. This explains the near absence of observed hollow-adsorbed CO on Pt(111) surfaces.

Certainly, there are other cases one can consider such as lower adsorbate coverage, and the topology of the top–\textit{hcp} hollow and hcp–fcc hollow diffusion barriers. With this example, however, we have demonstrated the efficacy of DMC 2-point extrapolations facilitated by error cancellation. These relatively inexpensive calculations can elucidate potential energy landscapes in surface chemistry with high accuracy.

\subsection{Estimation of Computational Gains}
\label{section:comp_gain}

Let us now estimate the computational savings achieved using the strategies described in the preceding discussions. Conventionally, DMC surface calculations report using three or four different supercell sizes, in some cases as large as 4$\times$4, or even 5$\times$5-tiled supercells of the minimal cell, to carry out finite-size extrapolations.\cite{shin_graphene,dmc_H2_on_Mg,dmc_water_on_LiH,hoggan_doblhoff-dier,Powell_JCP_2020} However, we have shown that error cancellation makes it numerically justifiable to perform 2-point extrapolations for binding energies, given reasonable agreement between the different possible extrapolation strategies (Fig. \ref{fig:internal_consistency_CO_Pt}). If the error cancellation is found to be particularly favorable, as seen in the case of H$_2$O/LiH(001) (Fig. \ref{fig:comparison_H2O_LiH}), one can also bypass subsequent finite-size extrapolations completely, relying on only a single, adequately large supercell for DMC binding energy calculations of different atomic configurations of interest. By reducing the sheer number of calculations required to extrapolate reliably to the thermodynamic limit, we effect \textit{at least linear} savings in computational cost.

To be conservative in our estimates of computational gain, let us suppose that one would ordinarily extrapolate using only a 2$\times$2 and a 3$\times$3 supercell, without accounting for error cancellation. We have verified, however, that using \textit{smaller} supercells for accurate binding energy extrapolation can be a valid strategy (Section \ref{section:2-point_extrapolation}). Hence, it may be reasonable to rely on the minimal 1$\times$1 cell instead of the 3$\times$3 supercell as an extrapolation data point (in addition to the 2$\times$2 supercell). This simple choice leads to a nine-fold reduction in the number of simulated electrons and a commensurate reduction in memory consumption as each DMC walker needs to track the trajectories of $9\times$ fewer electrons during statistical sampling. Considering that the time complexity of the DMC algorithm is $O(N^3)$ at best, this allows us a computational speedup of $\sim$700$\times$. This enables us to achieve appropriately small statistical error bars in a significantly shorter time.\cite{Foulkes_RMP_2001}


\section{Conclusions}

In this work, we have shown that, in Diffusion Monte Carlo surface calculations, the ubiquitous practice of taking energy differences between geometrically and chemically similar atomic configurations is accompanied by significant cancellation and accelerated convergence of many-body finite-size errors. We have exemplified these points by focusing on the examples of H$_2$O/LiH(001) and CO/Pt(111) adsorption. By assessing the quality of various single-supercell calculations and finite-size extrapolation strategies, we have demonstrated that one can rely on as few as one or two small supercell calculations for accurately predicting binding energies and potential energy surfaces in the thermodynamic limit. We have discussed how considerable computational gains may thus be achieved in QMC studies of catalysis by significantly reducing the simulation load associated with finite-size extrapolation or by bypassing the extrapolation entirely. These findings validate past researchers' use of these strategies to inexpensively evaluate thermodynamic binding energies\cite{Powell_JCP_2020,hoggan_doblhoff-dier,hsing_CO_on_Pt} and suggest that comparatively small DMC simulations of adsorption can yield meaningful insights into catalysis. 

In the context of the CO/Pt(111) system, the inexpensive 2-point extrapolation strategy was then used to accurately predict DMC binding energies and clearly establish the order of adsorption site preference as top $>$ bridge $>$ fcc hollow. We then used the same method to evaluate DMC energies for points on the top–fcc hollow diffusion trajectory suggested by nudged elastic band calculations. These energies distinctly revealed the existence of two potential energy barriers to the diffusion of CO from the one-fold (top) to the three-fold (fcc hollow) sites of Pt(111), providing a kinetic explanation for experimental observations of CO/Pt(111) adsorption. Furthermore, this validated the use of accurate and inexpensive extrapolation strategies facilitated by error cancellation in DMC calculations of surfaces.

We expect error cancellation to be a powerful tool for surface calculations when taking energy differences between geometrically and chemically similar atomic configurations. As seen in the examples in Sections \ref{section:h2o_lih} and \ref{section:co_pt}, such ``subtracted energies" are relevant in two important categories of energetic calculations: 1) \textit{Chemical modification}: Calculations of the energy associated with minor changes to chemical composition, such as the addition or removal of a small atomic/molecular adsorbate, impurity atom, vacancy or other point defect, etc.; and 2) \textit{Physical modification}: Mapping of potential energy landscapes associated with moderate rearrangement of atomic positions (w.r.t. some equilibrium configuration), such as the displacement of an adsorbate, diffusion of an impurity atom, vacancy or other point defect, axial straining of the structure, etc. While we have chosen to focus on adsorption here, preliminary results on sub-surface O-vacancy formation in TiO$_2$(110) indicate that error cancellation can be effectively utilized to reduce the cost of DMC calculations in these other classes of problems as well. The practical utility of error cancellation, and hence DMC methods, may thus be extended to a wide variety of problems in solid-state chemistry. We also expect this strategy to be applicable to three-dimensional bulk systems, not just surfaces. The only difference here would be in the use of an $O(1/N)$ extrapolation\cite{ceperley_bulk_scaling} instead of $O(N^{-5/4})$ in the two-dimensional case.

By highlighting how error cancellation can be leveraged to perform high-accuracy, low-cost DMC calculations in surface chemistry, we add to a growing body of efforts to identify and develop methods for accelerating highly accurate QMC calculations, including Jastrow sharing,\cite{ataca_jastrow_sharing} the hybrid orbital representation,\cite{hybridrep_luo} time-step error cancellation,\cite{size_consistency} and correlated sampling approaches,\cite{Shee_JCTC_2017,Hao_JPCL,Filippi_PRB_2000} to name a few. Our work also underscores how powerful combining DMC and embedding methods that rely on relatively small surface calculations may be for catalytic research.\cite{Petras_JCTC_2019} We believe this represents a step toward making high-throughput QMC methods more practically accessible to the broader computational catalysis research community.\\

\section*{Supporting Information}
Document includes binding energy benchmark comparisons, sample time-step convergence plot for CO/Pt(111), RMSD plot for H$_2$O/LiH(001) finite-size extrapolation, and top–hollow diffusion barrier plot using 2$\times$2 supercell and extrapolated DMC energies.

\acknowledgements
The authors thank Anouar Benali, Andrew Peterson, Andrew Medford, Shubham Sharma, Benjamin Comer, Richard Stratt, and J. Brad Marston for insightful discussions. G.R.I. was funded by AFOSR Award Number FA9550-19-1-9999, while B.M.R. graciously acknowledges funding from the U.S. Department of Energy, Office of Science, Basic Energy Sciences Award \#DE-FOA-0001912. This research was conducted using computational resources and services at the Center for Computation and Visualization, Brown University. 

\bibliography{ref}

\end{document}


\title{Supporting Information: Error Cancellation in Diffusion Monte Carlo Calculations of Surface Chemistry}

\author{Gopal R. Iyer}
\affiliation{Department of Chemistry, Brown University, Providence, RI 02912, USA}
\author{Brenda M. Rubenstein}
\affiliation{Department of Chemistry, Brown University, Providence, RI 02912, USA}

\begin{abstract}
\end{abstract}

\maketitle
\section{Benchmark Comparisons}
\subsection{\ce{CO/Pt}(111) Benchmark}
Hsing \textit{et al.}\cite{hsing_CO_on_Pt} performed DMC calculations of the binding energies of CO on Pt(111), among other similar systems. Since that paper only reports calculations for a single supercell size (2$\times$2 tiling of a $\sqrt{3}\times\sqrt{3}$ unit cell with three metal layers), we work with an identical geometry for comparison.

The DMC binding energy for top-adsorbed CO/Pt(111) reported in Ref. \citen{hsing_CO_on_Pt} is $-1.88\pm 0.065$ eV. As per the calculations in this work, the same binding energy for an identical geometry and tiling is found to be $-1.89\pm 0.052$ eV, which is in good agreement with the former.

However, as mentioned in the main text, Ref. \citen{hsing_CO_on_Pt} predicts that the top site is favored over the fcc hollow site by a much greater margin ($\sim$0.76 eV) than we do ($\sim$0.25 eV). This is the only other work we know of that uses DMC to study the CO/Pt(111) system. Previous DFT studies have routinely shown that differences in binding energies between these sites is rarely greater than 0.3 eV.\cite{Feibelman_JPCB_2001, Grinberg_JCP, Olsen_JCP, mason_prb, Janthon_JPCC} We take this to be a confirmation of the accuracy of our calculations.

\subsection{\ce{H2O/LiH(001)} Benchmark}
Tsatsoulis \textit{et al.}\cite{dmc_water_on_LiH} performed DMC calculations on the adsorption of a single \ce{H2O} molecule to a two-layer, 64-atom slab of LiH(001). A few notable departures in calculation details from this work include the height of the vacuum between periodic slab images in the vertical direction (20.5 Å vs. 15 Å in this work), the plane-wave energy cutoff used to generate the DFT trial wave function (500 eV vs. $\sim$4670 eV in this work), and the DMC timestep (0.02 a.u. vs. 0.01 a.u.) in this work. For more details, we refer the reader to Ref. \citen{dmc_water_on_LiH}.

Despite these differences in calculation details, the binding energy of \ce{H2O} on \ce{LiH}(001) extrapolated to infinite size is found to be $-275.69\pm 9.94$ meV which is within $\sim$1 mHa of the corresponding value reported in Ref. \citen{dmc_water_on_LiH} ($-250\pm 7$ meV). (We note that, in the original reference, a different sign convention is used, leading to positive binding energies.) Possible reasons for the slight discrepancies that appear in the extrapolated binding energies, in addition to the ones mentioned above, include the use of different pseudopotentials (Troullier–Martins correlation-consistent ECPs are used in this work), the use of the hybrid orbital representation in this work, and the use of different supercell sizes to carry out the finite-size extrapolation.\cite{hybridrep_luo}

\section{H$_2$O/L$\textrm{i}$H(001) RMSD Plot}
\begin{figure}
        \centering
         \includegraphics[width=9cm]{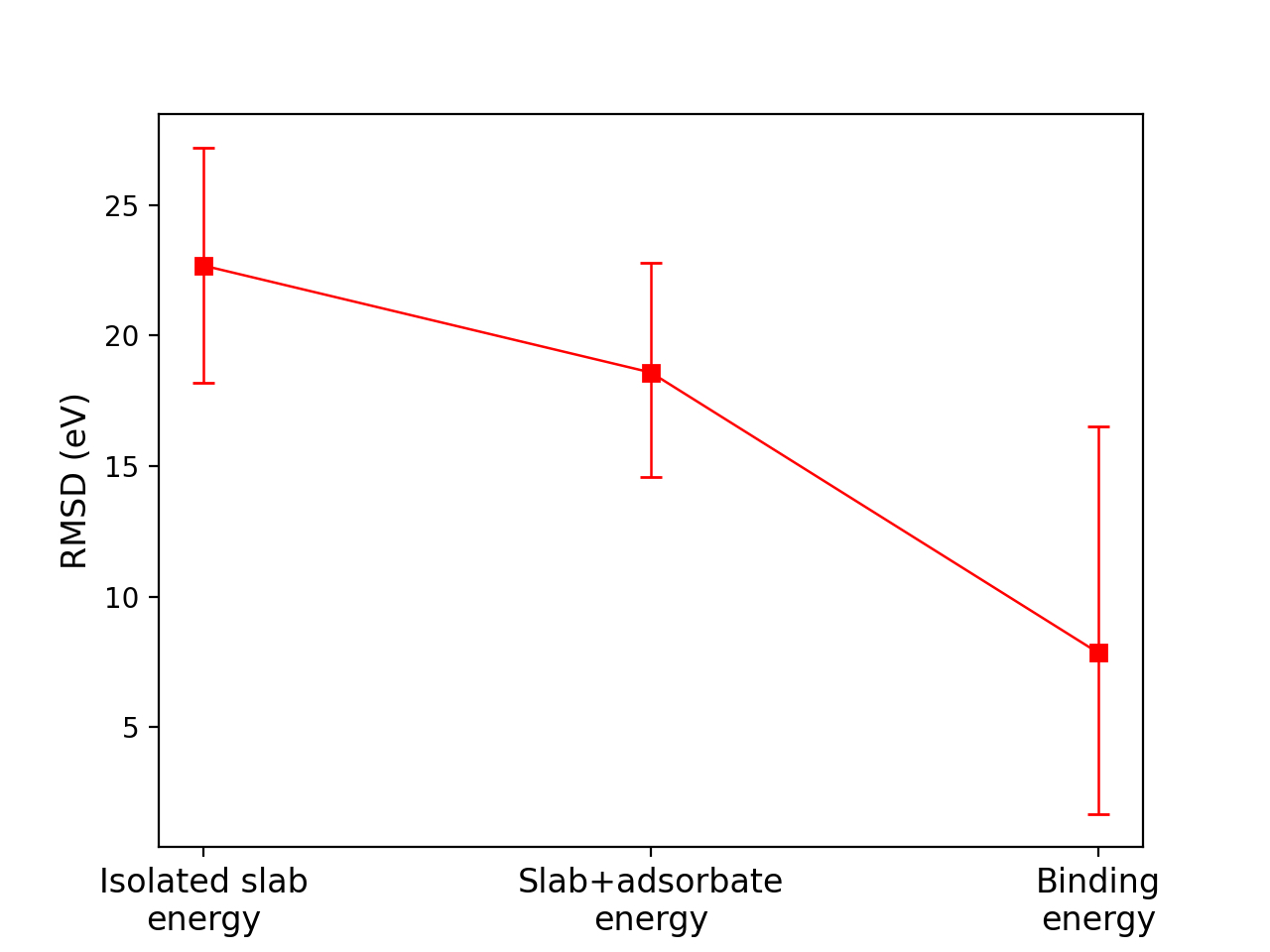}
         \caption{Comparison of RMSD in the finite-size extrapolation of absolute energies and additive interactions for \ce{H2O}/LiH(001).}
         \label{supplemental:rmsd_H2O_LiH}
\end{figure}

The RMSD of the extrapolation for the H$_2$O/LiH(001) system (Figs. \ref{supplemental:rmsd_H2O_LiH}) follows a very similar trend to that of the CO/Pt(111) system. The quality of the extrapolation for the binding energy is significantly better than for total energies of the isolated slab and slab/adsorbate systems.

\section{DMC Timestep Convergence}

\begin{figure}
        \centering
         \includegraphics[width=9cm]{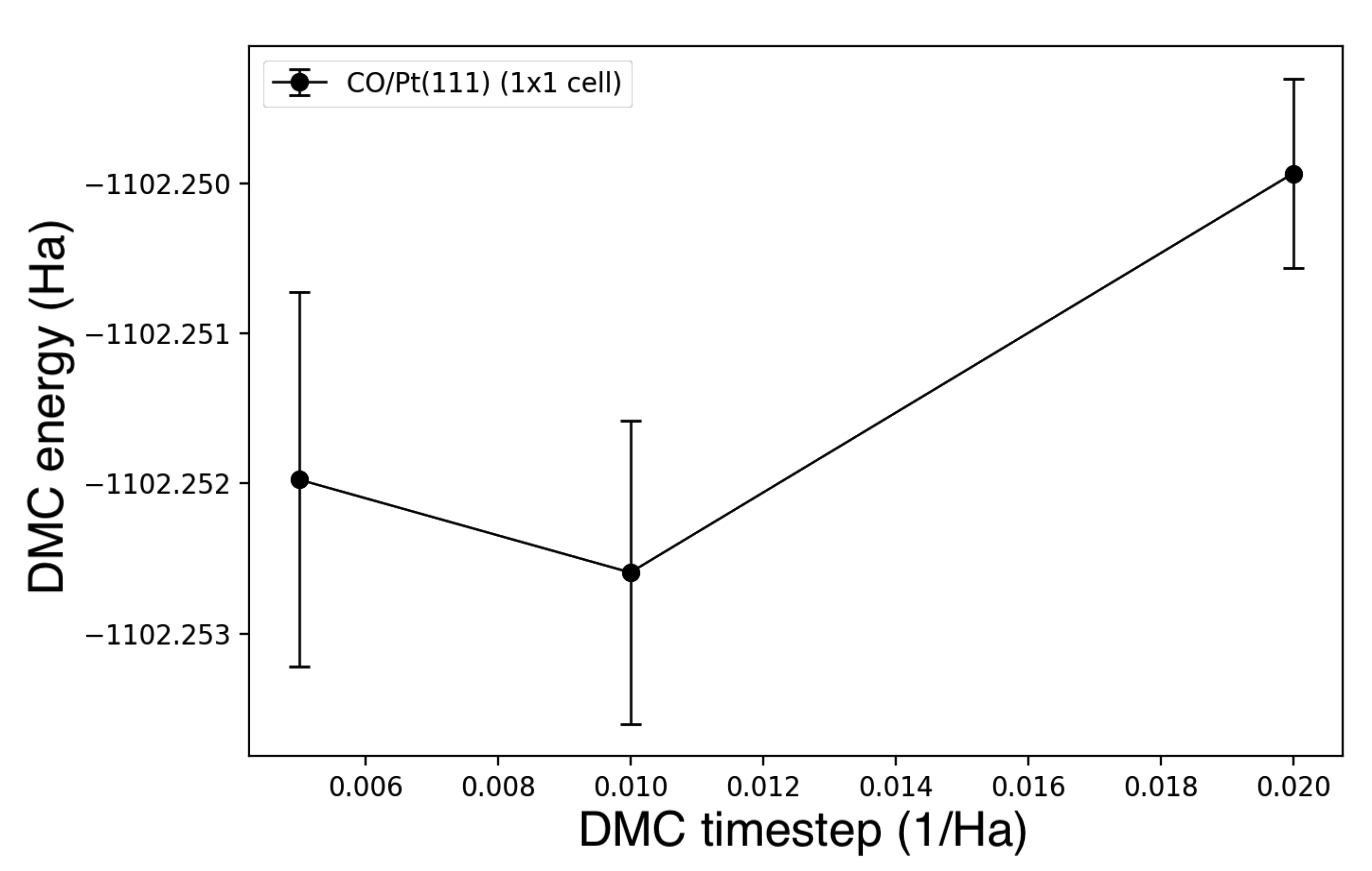}
         \caption{Time-step convergence of the CO/Pt(111) system for a 1x1 supercell suggests that a time-step of 0.01 Ha$^{-1}$ is a reasonable choice for DMC calculations.}
         \label{supplemental:timestep_CO_Pt}
\end{figure}

We have selected a DMC time-step of 0.01 Ha$^{-1}$ for each of our calculations. This is, in most cases, a reasonable choice of time-step for systems of the kind studied here.\cite{hsing_CO_on_Pt, Powell_JCP_2020} A representative time-step convergence plot for the CO/Pt(111) system is shown in Fig. \ref{supplemental:timestep_CO_Pt}. The difference between the energies obtained with time-step 0.01 Ha$^{-1}$ and time-step 0.005 Ha$^{-1}$ is within 1 mHa (ignoring statistical error).

\section{DMC Potential Energy Surface Using a Single Supercell}
\begin{figure}
        \centering
         \includegraphics[width=9cm]{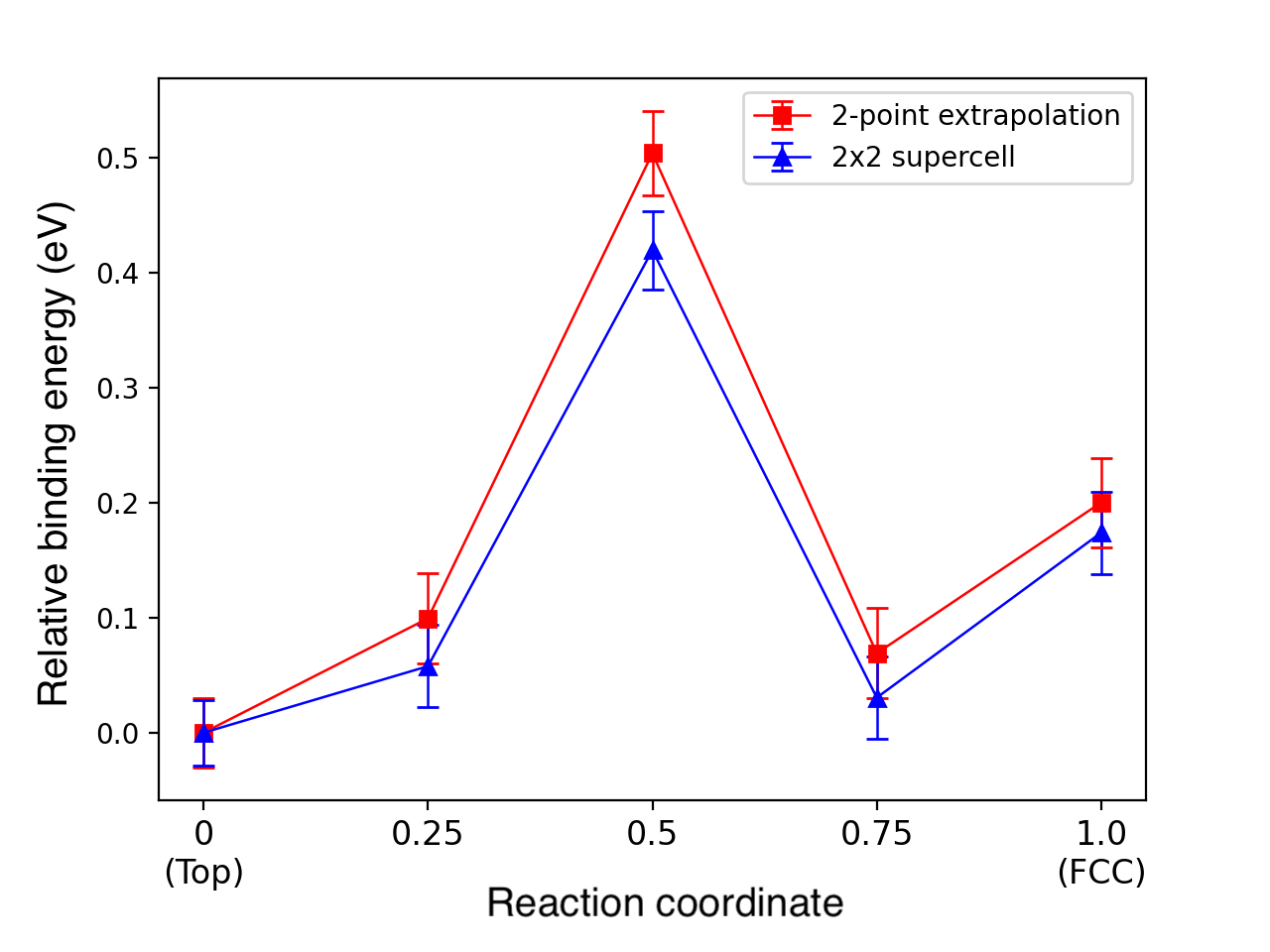}
         \caption{The top–fcc hollow diffusion barrier topology is largely captured using only a 2$\times$2 supercell compared to a 2-point extrapolation using 1$\times$1 and 2$\times$2 supercells.}
         \label{supplemental:single_supercell_pes}
\end{figure}
Fig. \ref{supplemental:single_supercell_pes} shows that using only a single 2$\times$2 supercell is largely sufficient to replicate the topology of the top–fcc hollow diffusion barrier. This suggests that the cancellation of many-body finite-size errors appears to show consistent behavior across different sites on the surface. Note that the value of the absolute binding energy of the top site predicted by the 2$\times$2 supercell alone does not lie within the region of agreement between different extrapolations shown in the main text.

\bibliography{ref}